\documentclass[12pt]{article}
\usepackage{graphicx}
\usepackage[bindingoffset=0cm,left=2cm,top=1.75cm,bottom=2cm,right=2.5cm]{geometry}

\begin{document}

\begin{center}
{\large \bf SPIN TOMOGRAPHY AND STAR--PRODUCT KERNEL \\ FOR QUBITS AND QUTRITS\\[10mm]}
{\bf Sergey N. Filippov$^{1}$ and Vladimir I. Man'ko$^{2}$}
\\[8mm]
\textit{$^{1}$ Moscow Institute of Physics and Technology (State University)\\
Institutskii per. 9, Dolgoprudnyi, Moscow Region 141700, Russia
\\[1mm]
$^{2}$ P. N. Lebedev Physical Institute, Russian Academy of Sciences\\
Leninskii Prospect 53, Moscow 119991, Russia
\\[1mm]}
e-mails: filippovsn@gmail.com \ manko@sci.lebedev.ru
\\[3mm]
\end{center}

\begin{abstract}
\noindent Using the irreducible tensor-operator technique, we
establish the relation between different forms of spin tomograms.
Quantizer and dequantizer operators are presented in simple
explicit forms and are specified for the low-spin states. The
kernel of the star-product is evaluated for qubits and qutrits and
its connection with a generic formula is found.
\end{abstract}

\textbf{Keywords:} spin tomography, quantizer, dequantizer,
star-product, kernel, qubit, qutrit.

\section{\label{introduction}Introduction}

According to the conventional treatment of quantum mechanics,
states of a system are associated either with the wave functions
(vectors in a Hilbert space) or with the density operators. Apart
from this, a new formulation of quantum states has been elaborated
\cite{tombesi-manko} (see also \cite{sudarshan-2008}) in the last
few decades. This representation associates the states with the
standard probability distributions. In fact, these probability
distributions can be measured directly in the experiment. All the
physical ingredients of quantum mechanics like means of
observables, their dispersions, etc. can be expressed in terms of
the probability distributions of the corresponding quantum states.
As far as continuous variables are concerned, the experiments to
reconstruct the Wigner function of photon states were performed,
for example in
\cite{raymer,mlynek,lvovsky,solimeno-porzio,bellini}. We point out
that in the approach \cite{tombesi-manko,sudarshan-2008} the
primary object in quantum mechanics associated to the quantum
states is namely the probability distribution. Once the
distribution is measured, it is not necessary at all to make any
intermediate steps (like reconstruction of Wigner function) in
order to extract the experimental information on the physical
properties of a system. This implies that the quantum properties
such as means of observables, variances, and other statistical
characteristics can be directly obtained in view of the
probability distributions. This aspect of the
probability-representation approach takes place also for the
states with discrete variables like spins, qubits, qudits, etc.

\medskip

We concentrate here on the problem of probability representation
for spin states. The quasidistribution functions for discrete
spin-variable states were discussed, for example, in
\cite{klimov-sanchez-soto,garcia-bondia}. The quasidistributions
such as analogs of the Wigner function \cite{wigner} or the Husimi
function \cite{husimi} for Lie groups, including $SU(2)$ group,
determine the corresponding states. In the same spirit, there
exists the possibility to use fair probability distributions for
spin degrees of freedom and, in fact, for other Lie groups (see,
e.g., \cite{klimov-manko}).

\medskip

Any spin state can be equivalently described by the density
operator $\hat{\rho}$ or by the fair probability-distribution
function called spin tomogram
\cite{dodonovPLA,oman'ko-jetp,weigert,amiet,d'ariano99,agarwal}
(for states with continuous variables see, e.g.,
\cite{tombesi-manko,sudarshan-2008,berber,vogel,mendes-physica,oman'ko-97,marmoJPA}).
The probability-distribution function is usually considered as an
intermediate procedure for the density operator reconstruction.
Apart from being a useful experimental tool, quantum tomograms
themselves are a primary notion of quantum states. Using
tomograms, one deals with functions instead of density operators.
Various properties of these functions are discussed in
\cite{andreev:manko,safonov:manko,andreev:safonov:manko,filipp}.
Similarly to the density operator, any other operator can be
identified with a function called tomographic symbol of the
operator. Unlike the tomogram, this function is not nonnegative in
the general case. To describe the standard product of operators on
a Hilbert space, one can introduce the star-product of the
tomographic symbols \cite{manko:star:1,manko:star:2}. The
star-product is associative but noncommutative in general.

\medskip

Operators and, in particular, observables can also be associated
with functions called dual tomographic symbols
\cite{omanko:vitale,vitale::dual} (the first step toward dual
symbols is taken in \cite{oman'ko-97}; dual symbols are applied to
study the quantumness of qubits in \cite{filip-preprint}).
Ordinary and dual tomographic symbols linked together enable one
to calculate expectation values of observables. Hence it is
possible to treat states, operators, and related quantities within
the framework of the unified tomographic representation.

\medskip

The aim of this paper is to reconsider quantizer and dequantizer
operators for spin tomograms of qudit states. These operators
relate tomograms with density operators, and observables with
ordinary and dual tomographic symbols as well. Moreover, quantizer
and dequantizer operators are constituent parts of the kernel of
star-product, which is widely used while dealing with maps of spin
operators onto functions. The general procedure to use quantizer
and dequantizer operators was discussed in the context of
star-product quantization schemes in
\cite{manko:star:2,omanko:vitale}. Although the explicit formulas
for quantizer and dequantizer operators were obtained earlier,
here we introduce another relatively simple form of these
operators, show simple relations between them, check the
equivalency of approaches applied in different works, and consider
cases of qubits and qutrits in detail. We also focus attention on
the star-product kernel for ordinary and dual tomographic symbols.

\medskip

The paper is organized as follows.

In Sec. \ref{ala:castanos}, we use the irreducible tensor-operator
technique to get the simple form of quantizer and dequantizer
operators. In Sec. \ref{section:xexponenta}, the exponential
representation of quantizer and dequantizer operators is
reconsidered in order to illustrate its equivalency to other
approaches. In Sec. \ref{section:delta}, we derive the kernel of
the unity operator on the set of qubit tomograms and that of on
the set of qutrit tomograms. In Sec. \ref{section:starproduct},
the explicit forms of the star-product kernel for qubits and
qutrits are obtained. In Sec. \ref{section:dual:symbols}, dual
tomographic symbols are briefly discussed. In Sec.
\ref{section:conclusions}, conclusions are presented.

\section{\label{ala:castanos} Irreducible
tensor-operator representation \\ of quantizer and dequantizer
operators}

Unless specifically stated, qudit states with spin $j$ are
considered. We start with state vectors $|jm\rangle$ and the
standard basis of the angular momentum operators $\hat{J}_{x}$,
$\hat{J}_{y}$, and $\hat{J}_{z}$ defined through

\begin{equation}
\hat{\bf J}^{2} |jm\rangle = j(j+1) |jm\rangle, \qquad \hat{J}_{z}
|jm\rangle = m |jm\rangle,
\end{equation}

\noindent where $m$ is the spin projection ($m=-j,-j+1,\dots,j$).

The spin tomogram of a qudit state given by its density operator
$\hat{\rho}$ reads

\begin{equation}
\label{tomogram} w({\bf x}) \equiv w(m,u) = \langle j m | u
\hat{\rho} u^{\dag} | j m \rangle = {\rm Tr} \Big( \hat{\rho}
u^{\dag} | j m \rangle \langle j m | u \Big) = {\rm Tr} \Big(
\hat{\rho} \hat{U}({\bf x}) \Big),
\end{equation}

\noindent where $u$ is a $(2j+1) \times (2j+1)$ unitary matrix of
irreducible representation of the rotation group $SU(2)$ and ${\bf
x}$ denotes the set of parameters $(m,u) \equiv
(m,\alpha,\beta,\gamma)$, with the Euler angles $\alpha$, $\beta$,
and $\gamma$ defining the matrix $u$.

The tomogram satisfies the following normalization conditions:

\begin{equation}
\sum \limits_{m=-j}^{j} w (m,u) = 1,  \qquad \frac{2j+1}{8\pi^{2}}
\int \limits_{0}^{2\pi} {\rm d}\alpha \int \limits_{0}^{\pi}
\sin\beta {\rm d}\beta \int \limits_{0}^{2\pi}{\rm d}\gamma \
w(m,\alpha,\beta,\gamma) = 1.
\end{equation}

We introduced the dequantizer operator in (\ref{tomogram}) as

\begin{equation}
\label{dequantizer:simple} \hat{U}({\bf x}) = u^{\dag} | j m
\rangle \langle j m | u
\end{equation}

\noindent which is nothing else but the spin-$j$ projector
operator onto the $m$ component along the $z$ axis rotated by an
element $u$ of $SU(2)$.

Given the tomogram $w({\bf x})$, one can reconstruct the density
operator $\hat{\rho}$ using the quantizer operator $\hat{D}({\bf
x})$ as follows

\begin{equation}
\label{rho:reconstruction} \hat{\rho} = \int w({\bf x})
\hat{D}({\bf x}) {\rm d}{\bf x},
\end{equation}

\noindent where

\begin{equation}
\int {\rm d}{\bf x} = \sum \limits_{m=-j}^{j} \frac{1}{8\pi^{2}}
\int \limits_{0}^{2\pi} {\rm d}\alpha \int \limits_{0}^{\pi}
\sin\beta {\rm d}\beta \int \limits_{0}^{2\pi}{\rm d}\gamma.
\end{equation}

Following \cite{castanos} we write the explicit formulas for both
dequantizer and quantizer operators in terms of the Clebsch-Gordan
coefficients $\langle j_{1}m_{1}; j_{2}m_{2} | j_{3}m_{3}
\rangle$:

\begin{eqnarray}
\label{dequantizer:complex} \hat{U}({\bf x}) =
\sum\limits_{L=0}^{2j} \sum\limits_{M=-L}^{L} (-1)^{j-m+M} \langle
jm;j-m | L0 \rangle D_{0
-M}^{(L)}(\alpha,\beta,\gamma) \hat{T}_{LM}^{(j)}, \\
\label{quantizer:complex} \hat{D}({\bf x}) =
\sum\limits_{L=0}^{2j} (2L+1) \sum\limits_{M=-L}^{L} (-1)^{j-m+M}
\langle jm;j-m | L0 \rangle D_{0 -M}^{(L)}(\alpha,\beta,\gamma)
\hat{T}_{LM}^{(j)},
\end{eqnarray}

\noindent where $D_{m_{1}m_{2}}^{(j)}(\alpha,\beta,\gamma)$ is the
Wigner $D$-function of the form

\begin{eqnarray}
\label{D-Wigner:function}
D_{m_{1}m_{2}}^{(j)}(\alpha,\beta,\gamma) = e^{-im_{2}\alpha}
e^{-im_{1}\gamma} \sum\limits_{s} &&
\frac{(-1)^{s}\sqrt{(j+m_{2})!(j-m_{2})!(j+m_{1})!(j-m_{1})!}}
{s!(j-m_{1}-s)!(j+m_{2}-s)!(m_{1}-m_{2}+s)!} \nonumber \\ &&
\times \left(\cos\frac{\beta}{2}\right)^{2j+m_{2}-m_{1}-2s}
\left(-\sin\frac{\beta}{2}\right)^{m_{1}-m_{2}+2s}
\end{eqnarray}

\noindent and $\hat{T}_{LM}^{(j)}$ is the irreducible tensor
operator for the $SU(2)$ group (also known as the polarization
operator \cite{varshalovich,klimov})

\begin{equation}
\hat{T}_{LM}^{(j)} = \sum \limits_{m_{1},m_{2}=-j}^{j}
(-1)^{j-m_{1}} \langle j m_{2}; j -m_{1} | LM \rangle | jm_{2}
\rangle \langle j m_{1} |.
\end{equation}

\noindent It is worth noting that the Clebsch-Gordan coefficients
can always be chosen real. Consequently, the operator
$\hat{T}_{LM}^{(j)}$ is real in the basis of states $|jm\rangle$.

From formula (\ref{dequantizer:simple}) it follows that, if the
operator $|jm\rangle\langle jm| = \hat{U}(m,0,0,0)$ is known, the
dequantizer can easily be calculated. So we focus on finding a
simple formula of this operator.

Since $D_{0 -M}^{(L)}(0,0,0) = \delta_{0 M}$, it follows from
(\ref{dequantizer:complex}) that

\begin{equation}
\label{jm:jm} |jm\rangle\langle jm| = \hat{U}(m,0,0,0) = \sum
\limits_{L=0}^{2j} (-1)^{j-m} \langle jm;j-m | L0 \rangle
\hat{T}_{L0}^{(j)} = \sum \limits_{L=0}^{2j} f_{L}^{(j)}(m)
\hat{S}_{L}^{(j)},
\end{equation}

\noindent where $f_{L}^{(j)}(m)$ is a function of the spin
projection $m$ and the operator $\hat{S}_{L}^{(j)}$ is
proportional to the operator $\hat{T}_{L0}^{(j)}$. Consequently,
$S_{L}^{(j)}$ is real and diagonal (and hence Hermitian) because
of peculiar form of the operator $\hat{T}_{L0}^{(j)}$

\begin{equation}
\label{TL0} \hat{T}_{L0}^{(j)} = \sum \limits_{m_{1}=-j}^{j}
(-1)^{j-m_{1}} \langle j m_{1}; j -m_{1} | L0 \rangle | jm_{1}
\rangle \langle j m_{1} |.
\end{equation}

Moreover, the operators $S_{L}^{(j)}$ and $S_{L'}^{(j)}$ are
orthogonal in the sense of trace operation

\begin{equation}
\label{S:S:orthogonality} {\rm Tr} \left(
\hat{S}_{L}^{(j)}\hat{S}_{L'}^{(j)} \right) \sim {\rm Tr} \left(
\hat{T}_{L0}^{(j)}\hat{T}_{L'0}^{(j)} \right) = \sum
\limits_{m=-j}^{j} \langle j m; j -m | L0 \rangle \langle j m; j
-m | L'0 \rangle = \delta_{L L'}.
\end{equation}

This implies that any Hermitian operator, being diagonal in the
basis of states $|jm\rangle$, can be resolved to the linear sum of
operators $\hat{S}_{L}^{(j)}$, $L=0,1,\dots,2j$. In other words,
matrices $S_{L}^{(j)}$ form a basis in the space of diagonal
Hermitian matrices. On the other hand, the operators
$\hat{J}_{z}^{k}$, $k=0,1,\dots,2j$ are also suitable to form the
basis in the same space of operators. The transition from one
basis to the other can be clarified by applying the operator
$\hat{P}$ which swaps states $|jm\rangle$ and $|j-m\rangle$.
Combining (\ref{TL0}) with such a rule, one obtains
$\hat{P}\hat{T}_{L0}^{(j)}\hat{P} = (-1)^{L}\hat{T}_{L0}^{(j)}$.
It is also obvious that $\hat{P}\hat{J}_{z}^{k}\hat{P} =
(-1)^{k}\hat{J}_{z}^{k}$. Hence, if the number $L$ is odd, the
operator $\hat{S}_{L}^{(j)}$ resolves to the sum of $\hat{J}_{z}$
to odd powers, and similarly, if the number $L$ is even, the
operator $\hat{S}_{L}^{(j)}$ resolves to the sum of $\hat{J}_{z}$
to even powers. Since $\hat{S}_{0}^{(j)} \sim \hat{J}_{z}^{0}$ and
$\hat{S}_{1}^{(j)} \sim \hat{J}_{z}^{1}$, we may assume that the
power of operators $\hat{J}_{z}$ in the expansion of
$\hat{S}_{L}^{(j)}$ is not greater than $L$. These results can be
summarized as follows:

\begin{eqnarray}
\label{S:even:through:J} & \hat{S}_{L}^{(j)} = \sum
\limits_{k=0}^{n} a_{2k}^{(j,L)} \hat{J}_{z}^{2k}
\qquad {\rm if} \qquad L=2n, \\
\label{S:odd:through:J} & \hat{S}_{L}^{(j)} = \sum
\limits_{k=0}^{n} b_{2k+1}^{(j,L)} \hat{J}_{z}^{2k+1} \qquad {\rm
if} \qquad L=2n+1,
\end{eqnarray}

\noindent or in the matrix form

\begin{equation}
\label{S:through:J::matrix}
\left(%
\begin{array}{c}
  \hat{S}_{0}^{(j)} \\
  \hat{S}_{1}^{(j)} \\
  \hat{S}_{2}^{(j)} \\
  \hat{S}_{3}^{(j)} \\
  \hat{S}_{4}^{(j)} \\
  \dots \\
\end{array}%
\right) = \left(%
\begin{array}{cccccc}
  a_{0}^{(j,0)} & 0 & 0 & 0 & 0 & \dots \\
  0 & b_{1}^{(j,1)} & 0 & 0 & 0 & \dots \\
  a_{0}^{(j,2)} & 0 & a_{2}^{(j,2)} & 0 & 0 & \dots \\
  0 & b_{1}^{(j,3)} & 0 & b_{3}^{(j,3)} & 0 & \dots \\
  a_{0}^{(j,4)} & 0 & a_{2}^{(j,4)} & 0 & a_{4}^{(j,4)} & \dots \\
  \dots & \dots & \dots & \dots & \dots & \dots \\
\end{array}%
\right) \left(%
\begin{array}{c}
  \hat{J}_{z}^{0} \\
  \hat{J}_{z}^{1} \\
  \hat{J}_{z}^{2} \\
  \hat{J}_{z}^{3} \\
  \hat{J}_{z}^{4} \\
  \dots \\
\end{array}%
\right).
\end{equation}

From (\ref{S:through:J::matrix}) it follows that ${\rm Tr}
(\hat{S}_{2n}^{(j)} \hat{S}_{2n+1}^{(j)}) = 0$, by construction.
The explicit form of coefficients $a_{2k}^{(j,L)}$ and
$b_{2k+1}^{(j,L)}$ can be found readily by employing the
orthogonality property (\ref{S:S:orthogonality}). In fact, since
the number of expansion terms in (\ref{S:even:through:J})
increases step-by-step with increase of $L$, any operator
$\hat{S}_{2n}^{(j)}$ must be orthogonal in the sense of trace
operation to each $\hat{J}_{z}^{2k}$, $k=0,1,\dots,n-1$. If we
combine this requirement with expansion (\ref{S:even:through:J}),
we get the following system of equations:

\begin{equation}
\label{a:coefficients::system}
\left(%
\begin{array}{cccc}
  {\rm Tr}\hat{J}_{z}^{0} & {\rm Tr}\hat{J}_{z}^{2} & \dots & {\rm Tr}\hat{J}_{z}^{2n-2} \\
  {\rm Tr}\hat{J}_{z}^{2} & {\rm Tr}\hat{J}_{z}^{4} & \dots & {\rm Tr}\hat{J}_{z}^{2n} \\
  \dots & \dots & \dots & \dots \\
  {\rm Tr}\hat{J}_{z}^{2n-2} & {\rm Tr}\hat{J}_{z}^{2n} & \dots & {\rm Tr}\hat{J}_{z}^{4n-4} \\
\end{array}%
\right) \left(%
\begin{array}{c}
  a_{0}^{(j,2n)} \\
  a_{2}^{(j,2n)} \\
  \dots \\
  a_{2n-2}^{(j,2n)} \\
\end{array}%
\right) = - a_{2n}^{(j,2n)} \left(%
\begin{array}{c}
  {\rm Tr}\hat{J}_{z}^{2n} \\
  {\rm Tr}\hat{J}_{z}^{2n+2} \\
  \dots \\
  {\rm Tr}\hat{J}_{z}^{4n-2} \\
\end{array}%
\right).
\end{equation}

\noindent It can be proved that the determinant $\Delta_{2n}$ of
square matrix in the left side of (\ref{a:coefficients::system})
is never equal to zero. This implies that one can calculate the
coefficients involved using the Cramer's rule \cite{korn-kramer}.
Indeed, let $a_{2n}^{(j,2n)}$ be equal to $-\Delta_{2n}$; then the
coefficients read

\begin{equation}
a_{2k}^{(j,2n)} = \Delta_{2n}^{(k+1)} \qquad {\rm if} \qquad
k=0,1,\dots,n-1, \qquad \qquad a_{2n}^{(j,2n)}=-\Delta_{2n},
\end{equation}

\noindent where $\Delta_{2n}^{(i)}$ is the determinant of the
matrix formed by replacing the $i$th column of matrix
(\ref{a:coefficients::system}) by the column vector $\left( {\rm
Tr}\hat{J}_{z}^{2n} \ {\rm Tr}\hat{J}_{z}^{2n+2} \ \dots \ {\rm
Tr}\hat{J}_{z}^{4n-2}\right)^{tr}$.

Arguing as above, we obtain

\begin{equation}
b_{2k+1}^{(j,2n+1)} = \Delta_{2n+1}^{(k+1)} \qquad {\rm if} \qquad
k=0,1,\dots,n-1, \qquad \qquad b_{2n+1}^{(j,2n+1)}=-\Delta_{2n+1},
\end{equation}

\noindent where

\begin{equation}
\label{matrix:TrJ::for:odd:b}
\Delta_{2n+1} = \det \left(%
\begin{array}{cccc}
  {\rm Tr}\hat{J}_{z}^{2} & {\rm Tr}\hat{J}_{z}^{4} & \dots & {\rm Tr}\hat{J}_{z}^{2n} \\
  {\rm Tr}\hat{J}_{z}^{4} & {\rm Tr}\hat{J}_{z}^{6} & \dots & {\rm Tr}\hat{J}_{z}^{2n+2} \\
  \dots & \dots & \dots & \dots \\
  {\rm Tr}\hat{J}_{z}^{2n} & {\rm Tr}\hat{J}_{z}^{2n+2} & \dots & {\rm Tr}\hat{J}_{z}^{4n-2} \\
\end{array}%
\right)
\end{equation}

\noindent and $\Delta_{2n+1}^{(i)}$ is the determinant of the
matrix formed by replacing the $i$th column of matrix
(\ref{matrix:TrJ::for:odd:b}) by the column vector $\left( {\rm
Tr}\hat{J}_{z}^{2n+2} \ {\rm Tr}\hat{J}_{z}^{2n+4} \ \dots \ {\rm
Tr}\hat{J}_{z}^{4n}\right)^{tr}$.

Though the explicit expressions for the coefficients
$a_{2k}^{(j,L)}$ and $b_{2k+1}^{(j,L)}$ seem rather complicated,
they can be readily computed by recalling that the spin projection
$m$ can take discrete values only. This results in the value of
${\rm Tr}\hat{J}_{z}^{k}$ being expressed by means of the
corresponding Bernoulli numbers \cite{korn-bernulli}.

Using formulas obtained, one can easily write the explicit form of
operators $\hat{S}_{L}^{(j)}$ in the case of small numbers $L$
(within a constant factor)

\begin{equation}
\hat{S}_{0}^{(j)} = \hat{J}_{z}^{0} = \hat{I}, \qquad
\hat{S}_{1}^{(j)} = \hat{J}_{z}, \qquad \hat{S}_{2}^{(j)} = 3
\hat{J}_{z}^{2} - j(j+1) \hat{I}, \qquad \hat{S}_{3}^{(j)} = 5
\hat{J}_{z}^{3} - (3j^{2}+3j-1)\hat{J}_{z}.
\end{equation}

Now we show how to calculate functions $f_{L}^{(j)}(m)$ which are
coefficients of expansion (\ref{jm:jm}). Using the orthogonality
property (\ref{S:S:orthogonality}), we obtain

\begin{equation}
\label{f:m::first} {\rm Tr} \left( \hat{S}_{L}^{(j)} | jm \rangle
\langle jm | \right) = \sum \limits_{L'=0}^{2j} f_{L'}^{(j)}(m)
{\rm Tr} \left( \hat{S}_{L}^{(j)} \hat{S}_{L'}^{(j)} \right) =
f_{L}^{(j)}(m) {\rm Tr} \left( \left. \hat{S}_{L}^{(j)} \right.
^{2} \right).
\end{equation}

\noindent On the other hand,

\begin{equation}
\label{f:m::second} {\rm Tr} \left( \hat{S}_{L}^{(j)} | jm \rangle
\langle jm | \right) = {\rm Tr} \left( \sum \limits_{k=0}^{L}
c_{k}^{(j,L)} \hat{J}_{z}^{k} | jm \rangle \langle jm | \right) =
\sum \limits_{k=0}^{L} c_{k}^{(j,L)} m^{k} {\rm Tr} \left( | jm
\rangle \langle jm | \right) = \sum \limits_{k=0}^{L}
c_{k}^{(j,L)} m^{k}.
\end{equation}

\noindent Combining (\ref{S:even:through:J}),
(\ref{S:odd:through:J}), (\ref{f:m::first}), and
(\ref{f:m::second}), we obtain

\begin{eqnarray}
\label{f:m:even} & f_{L}^{(j)}(m) = \left[ {\rm Tr} \left( \left.
\hat{S}_{L}^{(j)} \right. ^{2} \right) \right]^{-1} \sum
\limits_{k=0}^{n} a_{2k}^{(j,L)} m^{2k}
\qquad {\rm if} \qquad L=2n, \\
\label{f:m:odd} & f_{L}^{(j)}(m) = \left[ {\rm Tr} \left( \left.
\hat{S}_{L}^{(j)} \right. ^{2} \right) \right]^{-1} \sum
\limits_{k=0}^{n} b_{2k+1}^{(j,L)} m^{2k+1} \qquad {\rm if} \qquad
L=2n+1,
\end{eqnarray}

\noindent i.e., $f_{L}^{(j)}(m)$ has the same structure as the
operator $\hat{S}_{L}^{(j)}$. To be more precise, one should
simply replace the operator $\hat{J}_{z}$ by the variable $m$ and
divide the result by the normalization coefficient.

Using (\ref{TL0}) it is not hard to prove that the functions
$f_{L}^{(j)}(m)$ are expressed by means of the Clebsch-Gordan
coefficients as follows:

\begin{equation}
f_{L}^{(j)}(m) = \left[ {\rm Tr} \left( \left. \hat{S}_{L}^{(j)}
\right. ^{2} \right) \right]^{-1/2} (-1)^{j-m} \langle jm ; j-m |
L0 \rangle.
\end{equation}

\noindent Employing the known properties of the Clebsch-Gordan
coefficients \cite{varshalovich} leads to a recurrence relation of
the form

\begin{eqnarray}
&& \!\!\! f_{L}^{(j)}(m) = \left[
\frac{4(2L-1)(2L+1)}{L^{2}(2j-L+1)(2j+L+1){\rm Tr} \Big( \left.
\hat{S}_{L}^{(j)} \right. ^{2}\Big)} \right]^{1/2} \\
&& \!\!\! \times \left\{ \left[ {\rm Tr} \Big( \left.
\hat{S}_{L-1}^{(j)} \right. ^{2} \Big) \right]^{1/2} m
f_{L-1}^{(j)}(m) - \left[ \frac{(L-1)^{2}(2j-L+2)(2j+L) {\rm Tr}
\Big( \left. \hat{S}_{L-2}^{(j)} \right. ^{2}\Big)}{4(2L-3)(2L-1)}
\right]^{1/2} f_{L-2}^{(j)}(m) \right\}. \nonumber
\end{eqnarray}

\bigskip

Let us illustrate the results obtained by examples.

\noindent {\bf Qubit}

\begin{equation}
| 1/2, m \rangle \langle 1/2, m | = \frac{1}{2}
\hat{I} + 2m \hat{J}_{z} = \frac{1}{2} \left(%
\begin{array}{cc}
  1 & 0 \\
  0 & 1 \\
\end{array}%
\right) + m \left(%
\begin{array}{cc}
  1 & 0 \\
  0 & -1 \\
\end{array}%
\right)
\end{equation}

\noindent {\bf Qutrit}

\begin{eqnarray}
| 1, m \rangle \langle 1, m | && = \frac{1}{3} \hat{I} +
\frac{m}{2} \hat{J}_{z} + \frac{3m^{2}-2}{6} \left(
3\hat{J}_{z}^{2} - 2
\hat{I} \right) \nonumber \\
&& = \frac{1}{3} \left(%
\begin{array}{ccc}
  1 & 0 & 0 \\
  0 & 1 & 0 \\
  0 & 0 & 1 \\
\end{array}%
\right) + \frac{m}{2} \left(%
\begin{array}{ccc}
  1 & 0 & 0 \\
  0 & 0 & 0 \\
  0 & 0 & -1 \\
\end{array}%
\right) + \frac{3m^{2}-2}{6} \left(%
\begin{array}{ccc}
  1 & 0 & 0 \\
  0 & -2 & 0 \\
  0 & 0 & 1 \\
\end{array}%
\right).
\end{eqnarray}

\noindent {\bf Qudit with spin $j=3/2$}

\begin{eqnarray}
| 3/2, m \rangle \langle 3/2, m | && = \frac{1}{4} \hat{I} +
\frac{m}{5} \hat{J}_{z} + \frac{4m^{2}-5}{64} \left(
4\hat{J}_{z}^{2} - 5\hat{I} \right) + \frac{20m^{3}-41m}{720}
\left(
20\hat{J}_{z}^{3} - 41\hat{J}_{z} \right) \nonumber \\
&& = \frac{1}{4} \left(%
\begin{array}{cccc}
  1 & 0 & 0 & 0 \\
  0 & 1 & 0 & 0 \\
  0 & 0 & 1 & 0 \\
  0 & 0 & 0 & 1 \\
\end{array}%
\right) + \frac{m}{10} \left(%
\begin{array}{cccc}
  3 & 0 & 0 & 0 \\
  0 & 1 & 0 & 0 \\
  0 & 0 & -1 & 0 \\
  0 & 0 & 0 & -3 \\
\end{array}%
\right)  \nonumber \\
&&  + \frac{4m^{2}-5}{16} \left(%
\begin{array}{cccc}
  1 & 0 & 0 & 0 \\
  0 & -1 & 0 & 0 \\
  0 & 0 & -1 & 0 \\
  0 & 0 & 0 & 1 \\
\end{array}%
\right) + \frac{20m^{3}-41m}{120} \left(%
\begin{array}{cccc}
  1 & 0 & 0 & 0 \\
  0 & -3 & 0 & 0 \\
  0 & 0 & 3 & 0 \\
  0 & 0 & 0 & -1 \\
\end{array}%
\right).
\end{eqnarray}

Now, in view of the explicit form of expansion (\ref{jm:jm}),
recalling (\ref{dequantizer:simple}), we obtain the following
formula for the dequantizer operator:

\begin{equation}
\label{dequantizer:expansion:S} \hat{U}({\bf x}) = \sum
\limits_{L=0}^{2j} f_{L}^{(j)}(m) \ u^{\dag} \hat{S}_{L}^{(j)} u.
\end{equation}

A comparison of (\ref{dequantizer:complex}) with
(\ref{quantizer:complex}) leads to a simple form of the quantizer
operator

\begin{equation}
\label{quantizer:expansion:S} \hat{D}({\bf x}) = \sum
\limits_{L=0}^{2j} (2L+1) f_{L}^{(j)}(m) \ u^{\dag}
\hat{S}_{L}^{(j)} u.
\end{equation}

Using these formulas together with the examples considered above,
we write the dequantizer operator for qubits

\begin{equation}
\label{U:qubit:explicitmatrix} \hat{U}({\bf x}) = \frac{1}{2} \left(%
\begin{array}{cc}
  1 & 0 \\
  0 & 1 \\
\end{array}%
\right) + m \left(%
\begin{array}{cc}
  \cos\beta & -e^{i\alpha}\sin\beta \\
  -e^{-i\alpha}\sin\beta & -\cos\beta \\
\end{array}%
\right)
\end{equation}

\noindent  and for qutrits

\begin{eqnarray}
\label{U:qutrit:explicitmatrix}  \hat{U}({\bf x})  = && \frac{1}{3} \left(%
\begin{array}{ccc}
  1 & 0 & 0 \\
  0 & 1 & 0 \\
  0 & 0 & 1 \\
\end{array}%
\right) + \frac{m}{2} \left(%
\begin{array}{ccc}
  \cos\beta & -\displaystyle{\frac{\sin\beta}{\sqrt{2}}\ e^{i\alpha}} & 0 \\
  -\displaystyle{\frac{\sin\beta}{\sqrt{2}}\ e^{-i\alpha}} & 0 & -\displaystyle{\frac{\sin\beta}{\sqrt{2}}\ e^{i\alpha}} \\
  0 & -\displaystyle{\frac{\sin\beta}{\sqrt{2}}\ e^{-i\alpha}} & -\cos\beta \\
\end{array}%
\right)  \nonumber\\
&& + \frac{3m^{2}-2}{6} \left(%
\begin{array}{ccc}
  \displaystyle{\frac{3\cos^{2}\beta -1}{2}} & -\displaystyle{\frac{3\cos\beta\sin\beta}{\sqrt{2}}\ e^{i\alpha}} & \displaystyle{\frac{3\sin^{2}\beta}{2}\ e^{i2\alpha}} \\
  -\frac{3\cos\beta\sin\beta}{\sqrt{2}}\ e^{-i\alpha} & -\left(3\cos^{2}\beta -1\right) & \displaystyle{\frac{3\cos\beta\sin\beta}{\sqrt{2}}\ e^{i\alpha}} \\
  \displaystyle{\frac{3\sin^{2}\beta}{2}}\ e^{-i2\alpha} & \displaystyle{\frac{3\cos\beta\sin\beta}{\sqrt{2}}\ e^{-i\alpha}} & \displaystyle{\frac{3\cos^{2}\beta -1}{2}} \\
\end{array}%
\right).
\end{eqnarray}

The quantizer operator for qubits is merely obtained from
(\ref{U:qubit:explicitmatrix}) by multiplying the second term by
3. The quantizer operator for qutrits is obtained from
(\ref{U:qutrit:explicitmatrix}) by multiplying the second and
third terms by 3 and 5, respectively. It is worth noting that the
quantzer and dequantizer operators of spin states are Hermitian
[see (\ref{dequantizer:expansion:S}) and
(\ref{quantizer:expansion:S})]. In addition, the dequantizer is
positive as well. The other remarkable fact for both dequantizer
and quantizer operators is that the matrix elements of
$u^{\dag}{S}_{L}^{(j)}u$ are in close relation to associated
Legendre functions of degree $L$ and different orders proportional
to the distance to the leading diagonal. This is also the argument
for the spin tomogram $w(m,\alpha,\beta)$ (independent on
$\gamma$) to be a finite sum of spherical functions
$Y_{l}^{m}(\beta,\alpha)$, $l=0,1,\dots,2j$; this fact has been
emphasized earlier in \cite{andreev:manko}.


Low-spin tomograms are of particular interest here because any
qudit tomogram and the photon-number tomogram with infinite
outputs can be mapped onto qubit or qutrit tomogram
\cite{chernega,filipp-qubit-portrait}.

\begin{figure}
\begin{center}
\includegraphics{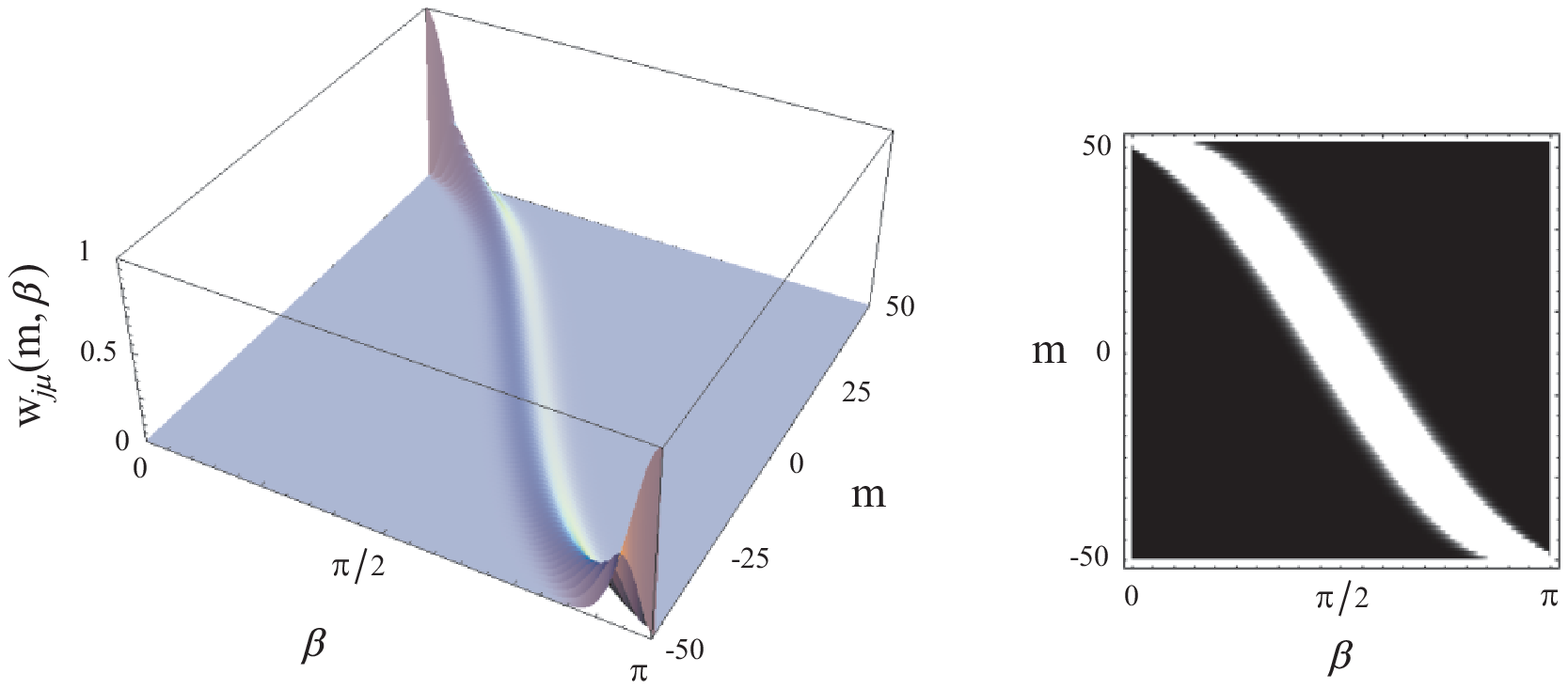}
\includegraphics{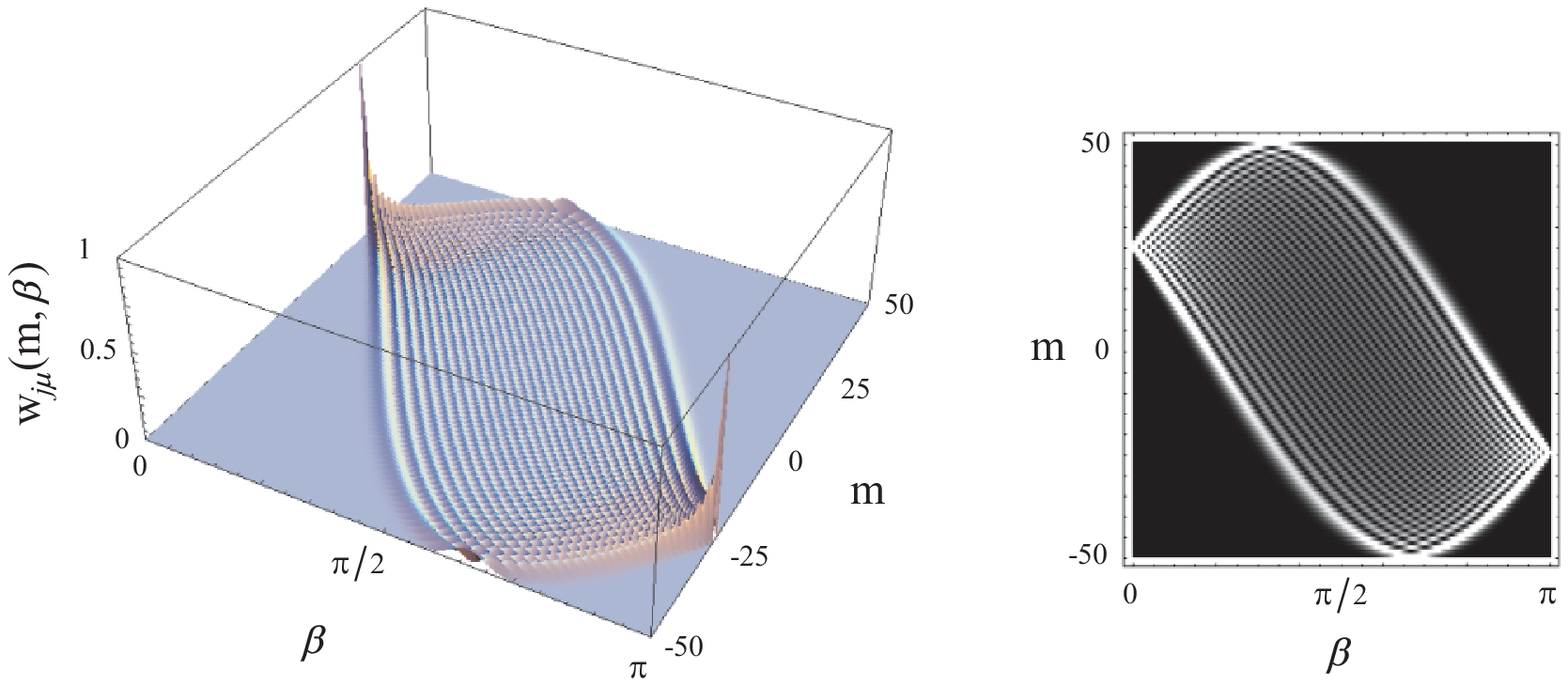}
\includegraphics{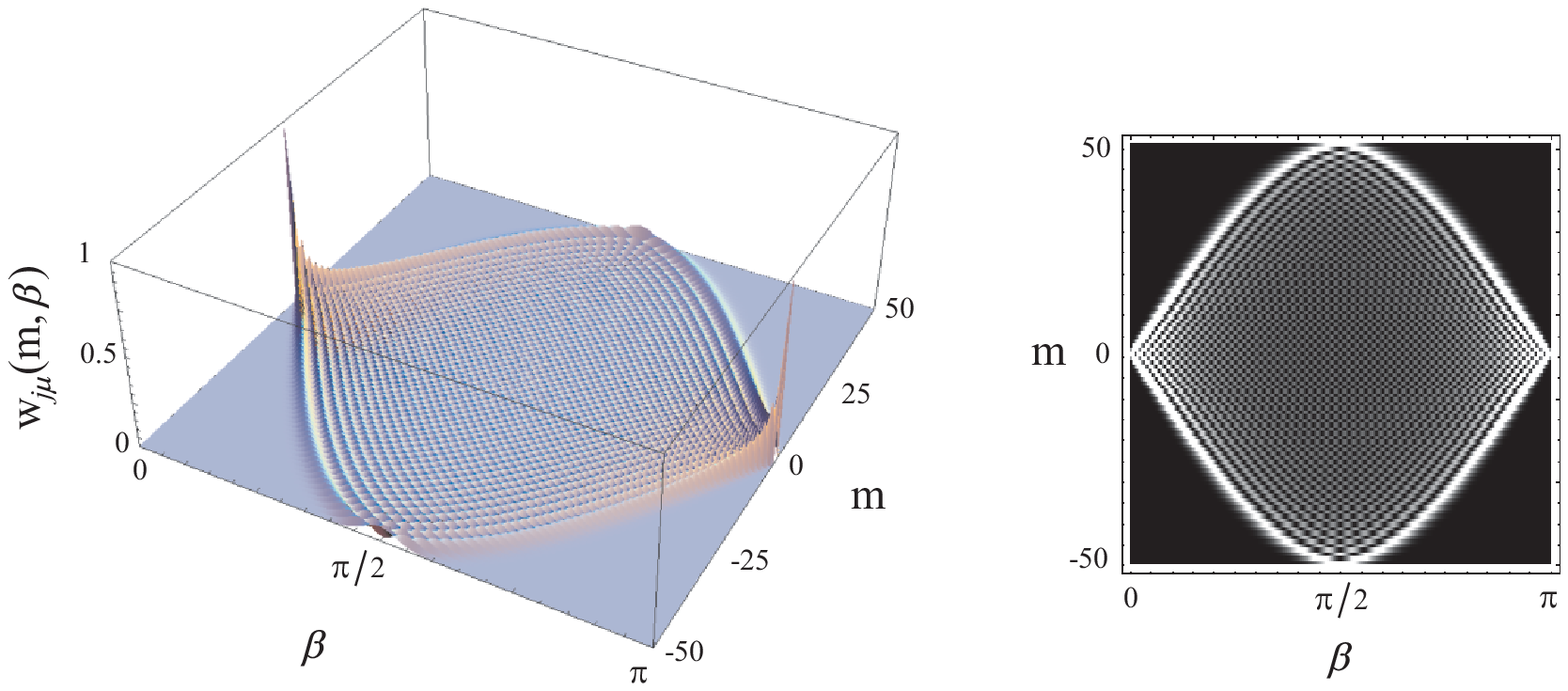}
\caption{\label{graph-50} Spin tomograms (on the left) and density
plots (on the right) of the state $|j\mu\rangle$ with $j=50$ and
$\mu=50$ (top), $j=50$ and $\mu=25$ (middle), and $j=50$ and
$\mu=0$ (bottom).}
\end{center}
\end{figure}

The quasiprobability-distribution functions of continuous
variables are usually illustrated by plotting on the corresponding
phase space. Here, we give an illustration of the spin tomogram
$w_{j\mu}({\bf x})$ of the pure state $|j \mu\rangle$. The
tomogram reads

\begin{equation}
\label{pure:state:tomogram} w_{j\mu}({\bf x}) = {\rm Tr} \Big( |j
\mu\rangle \langle j \mu | \hat{U}({\bf x}) \Big)= \left| \langle
j m | u |j \mu\rangle \right|^{2} = \left|
D_{m\mu}^{(j)}(\alpha,\beta,\gamma) \right|^{2}
\end{equation}

\noindent where the function $D_{m \mu}^{(j)}(\alpha, \beta,
\gamma)$ is given by (\ref{D-Wigner:function}). From this follows
that the tomogram depends only on the Euler angle $\beta$, i.e.,
$w_{j \mu}({\bf x}) = w_{j\mu}(m,\beta)$. The different examples
of this tomogram are depicted in Fig. \ref{graph-50}. It is worth
noting that the tomogram (\ref{pure:state:tomogram}) tends to the
following asymptotic function as $j \rightarrow \infty$
\cite{castanos}:

\begin{equation}
\label{pure:tomogram:limit} \tilde{w}_{j\mu}(m, \beta) = \left(
\pi j \sin^{2}\beta \right)^{-1/2} \left[ 2^{j-\mu} (j-\mu)!
\right]^{-1} \exp \left( -2j\sin^{2}\beta \right) H_{j-\mu}^{2}
\left( \frac{m-j\cos\beta}{\sqrt{j}\sin\beta} \right),
\end{equation}

\noindent where $H_{n}(x)$ is the Hermite polynomial of degree
$n$. The strong dependence of the asymptotic function
(\ref{pure:tomogram:limit}) on the value of $\beta$ occurs due to
a pointwise but nonuniform convergence of
(\ref{pure:state:tomogram}) to (\ref{pure:tomogram:limit}).

\section{\label{section:xexponenta} Exponential representation \\ of quantizer and dequantizer operators}

The dequantizer operator can be alternatively expressed in terms
of the Kronecker delta-symbol \cite{mmanko}, i.e., in the form of
the following exponential operator:

\begin{equation}
\label{dequantizer:exp} \hat{U}({\bf x}) = \delta \left( m -
u^{\dag}\hat{J}_{z}u \right) = \frac{1}{2\pi} \int
\limits_{0}^{2\pi} \exp \left[ i\left( m - u^{\dag}\hat{J}_{z}u
\right) \varphi \right] {\rm d} \varphi.
\end{equation}

\noindent Let us check that such a representation of the
dequantizer operator completely coincides with that discussed in
previous section. To start, we notice that the operator
$u\hat{U}({\bf x})u^{\dag}$ is diagonal in the basis of states
$|jm\rangle$. Consequently, it can be resolved to the linear sum
of operators $\hat{S}_{L}^{(j)}$. Indeed,

\begin{equation}
u\hat{U}({\bf x})u^{\dag} = \frac{1}{2\pi} \int \limits_{0}^{2\pi}
\exp \left[ i\left( m - \hat{J}_{z} \right) \varphi \right] {\rm
d} \varphi = \sum \limits_{L=0}^{2j} g_{L}^{(j)}(m)
\hat{S}_{L}^{(j)},
\end{equation}

\noindent where the expansion coefficients read

\begin{equation}
\label{g::coefficient:for:exponential:expansion} g_{L}^{(j)}(m) =
\left[ {\rm Tr} \left( \left. \hat{S}_{L}^{(j)} \right. ^{2}
\right) \right]^{-1} \frac{1}{2\pi} \int \limits_{0}^{2\pi} {\rm
Tr} \left\{ \exp \left[ i\left( m - \hat{J}_{z} \right) \varphi
\right] \hat{S}_{L}^{(j)} \right\} {\rm d} \varphi.
\end{equation}

\noindent Now, in view of (\ref{S:even:through:J}) and
(\ref{S:odd:through:J}) or, in general, $\hat{S}_{L}^{(j)} =
\sum_{k=0}^{L} c_{k}^{(j,L)} \hat{J}_{z}^{k}$, we obtain

\begin{equation}
{\rm Tr} \left\{ \exp \left[ i\left( m - \hat{J}_{z} \right)
\varphi \right] \hat{S}_{L}^{(j)} \right\} = \sum
\limits_{m'=-j}^{j} e^{i(m-m')\varphi} \sum_{k=0}^{L}
c_{k}^{(j,L)} m'^{k}.
\end{equation}

\noindent and (\ref{g::coefficient:for:exponential:expansion})
takes the form

\begin{equation}
\label{g::coeff:consequence} g_{L}^{(j)}(m) = \left[ {\rm Tr}
\left( \left. \hat{S}_{L}^{(j)} \right. ^{2} \right) \right]^{-1}
\sum \limits_{m'=-j}^{j} \sum_{k=0}^{L} c_{k}^{(j,L)} m'^{k}
\frac{1}{2\pi} \int \limits_{0}^{2\pi} e^{i(m-m')\varphi}  {\rm d}
\varphi = \left[ {\rm Tr} \left( \left. \hat{S}_{L}^{(j)} \right.
^{2} \right) \right]^{-1} \sum_{k=0}^{L} c_{k}^{(j,L)} m^{k}.
\end{equation}

\noindent Comparing (\ref{g::coeff:consequence}) with
(\ref{f:m:even}) and (\ref{f:m:odd}), we conclude that
$g_{L}^{(j)}(m) \equiv f_{L}^{(j)}(m)$. Therefore, we have proved
that as far as the dequantizer operator is concerned, exponential
expression (\ref{dequantizer:exp}) is equivalent to expansion
(\ref{dequantizer:expansion:S}) through orthogonal operators
$\hat{S}_{L}^{(j)}$ and, consequently, to formula
(\ref{dequantizer:complex}) expressed in terms of irreducible
tensor operators.

Similarly to the case of dequantizer, the quantizer operator
$\hat{D}({\bf x})$ can be represented in the exponential form
\cite{d'ariano:paini}

\begin{eqnarray}
\label{quantizer:exp} \hat{D}({\bf x}) && = \frac{2j+1}{\pi} \int
\limits_{0}^{2\pi} \sin^{2} \frac{\varphi}{2} \exp \left[ i\left(
m - u^{\dag}\hat{J}_{z}u \right) \varphi \right] {\rm d} \varphi
\nonumber\\
&& = u^{\dag} \left\{ \frac{2j+1}{\pi} \int \limits_{0}^{2\pi}
\sin^{2} \frac{\varphi}{2} \exp [ i( m - \hat{J}_{z} ) \varphi ]
{\rm d} \varphi \right\} u.
\end{eqnarray}

\noindent Let us consider the operator $u\hat{D}({\bf x})u^{\dag}$
in detail. In fact, it follows easily that

\begin{equation}
\label{uDu::exp} u\hat{D}({\bf x})u^{\dag} = \frac{2j+1}{2\pi}
\int \limits_{0}^{2\pi} \left\{ e^{ i( m - \hat{J}_{z} ) \varphi }
- \frac{1}{2} e^{ i( m + 1 - \hat{J}_{z} ) \varphi } - \frac{1}{2}
e^{ i( m - 1 - \hat{J}_{z} ) \varphi }  \right\} {\rm d} \varphi.
\end{equation}

\noindent Since the spin projection to an arbitrary axis can only
take values from $-j$ to $j$, we obtain

\begin{eqnarray}
&& \frac{1}{2\pi} \int \limits_{0}^{2\pi} e^{ i( m - \hat{J}_{z} )
\varphi } {\rm d} \varphi = |jm\rangle \langle jm| = u\hat{U}({\bf x})u^{\dag}, \\
&& \frac{1}{2\pi} \int \limits_{0}^{2\pi} e^{ i( m + 1 -
\hat{J}_{z} ) \varphi } {\rm d} \varphi = \hat{R}_{+} |jm\rangle
\langle jm| \hat{R}_{-} = \left\{
\begin{array}{l}
  |j,m+1\rangle \langle j, m+1| \ \ {\rm if} \ \  m = -j,\dots,j-1, \\
  0 \ \ {\rm if} \ \  m = j, \\
\end{array} \right. \ \ \ \  \\
&& \frac{1}{2\pi} \int \limits_{0}^{2\pi} e^{ i( m - 1 -
\hat{J}_{z} ) \varphi } {\rm d} \varphi = \hat{R}_{-} |jm\rangle
\langle jm| \hat{R}_{+} = \left\{ \begin{array}{l}
  |j,m-1\rangle \langle j, m-1| \ \ {\rm if} \ \  m = -j+1,\dots,j, \\
  0 \ \ {\rm if} \ \  m = -j, \\
\end{array} \right. \label{m-1-J::exp}
\end{eqnarray}

\noindent where we introduced operators $\hat{R}_{+}$ and
$\hat{R}_{-}$ specified by their matrices in the basis of states
$|jm\rangle$

\begin{equation}
\label{R+R-::matrices}
R_{+} = \left(%
\begin{array}{cccccc}
  0 & 1 & 0 & \dots & 0 & 0 \\
  0 & 0 & 1 & \dots & 0 & 0 \\
  0 & 0 & 0 & \dots & 0 & 0 \\
  \dots & \dots & \dots & \dots & \dots & \dots \\
  0 & 0 & 0 & \dots & 0 & 1 \\
  0 & 0 & 0 & \dots & 0 & 0 \\
\end{array}%
\right), \qquad R_{-} = R_{+}^{\dag} = \left(%
\begin{array}{cccccc}
  0 & 0 & 0 & \dots & 0 & 0 \\
  1 & 0 & 0 & \dots & 0 & 0 \\
  0 & 1 & 0 & \dots & 0 & 0 \\
  \dots & \dots & \dots & \dots & \dots & \dots \\
  0 & 0 & 0 & \dots & 0 & 0 \\
  0 & 0 & 0 & \dots & 1 & 0 \\
\end{array}%
\right).
\end{equation}

\noindent Combining (\ref{uDu::exp})$-$(\ref{m-1-J::exp}), we
obtain the explicit relation between quantizer and dequantizer
operators

\begin{equation}
\label{D:through:U} \hat{D}({\bf x}) \equiv \hat{D}(m,u) = (2j+1)
\left[ \hat{U}({\bf x}) - \frac{1}{2} \hat{R}_{+}(u) \hat{U}({\bf
x}) \hat{R}_{-}(u) - \frac{1}{2} \hat{R}_{-}(u) \hat{U}({\bf x})
\hat{R}_{+}(u) \right],
\end{equation}

\noindent where $\hat{R}_{+}(u)=u^{\dag}\hat{R}_{+}u$ and
$\hat{R}_{-}(u)=u^{\dag}\hat{R}_{-}u$. It is easy to prove that
the inverse formula reads

\begin{equation}
\label{U:through:D} \hat{U}({\bf x}) = \frac{1}{2j+1} \sum
\limits_{k=0}^{\infty} \hat{D}^{(k)},
\end{equation}

\noindent with $\hat{D}^{(k)}$ being defined by the recurrence
relations

\begin{equation}
\hat{D}^{(k)} = \frac{1}{2} \left[ \hat{R}_{+}(u) \hat{D}^{(k-1)}
\hat{R}_{-}(u) + \hat{R}_{-}(u) \hat{D}^{(k-1)}
\hat{R}_{+}(u)\right], \qquad \hat{D}^{(0)} = \hat{D}({\bf x}).
\end{equation}

Now let us show that the exponential form of the quantizer
operator (\ref{quantizer:exp}) is in complete agreement with
formula (\ref{quantizer:expansion:S}). Using the explicit
expression (\ref{D:through:U}) of quantizer operator through the
dequantizer, employing expansion (\ref{dequantizer:expansion:S}),
proved to be identical to the exponential form, we arrive at

\begin{equation}
\label{uDu::complex:S} u\hat{D}({\bf x})u^{\dag} = (2j+1)
\sum\limits_{L=0}^{2j} f_{L}^{(j)}(m) \left( \hat{S}_{L}^{(j)} -
\frac{1}{2} \hat{R}_{+} \hat{S}_{L}^{(j)} \hat{R}_{-} -
\frac{1}{2} \hat{R}_{-} \hat{S}_{L}^{(j)} \hat{R}_{+}\right).
\end{equation}

\noindent On the other hand, the diagonal operator $\left(
\hat{S}_{L}^{(j)} - \frac{1}{2} \hat{R}_{+} \hat{S}_{L}^{(j)}
\hat{R}_{-} - \frac{1}{2} \hat{R}_{-} \hat{S}_{L}^{(j)}
\hat{R}_{+}\right)$ can also be resolved to the sum

\begin{equation}
\label{S-RSR-RSR::through:h:S} \hat{S}_{L}^{(j)} - \frac{1}{2}
\hat{R}_{+} \hat{S}_{L}^{(j)} \hat{R}_{-} - \frac{1}{2}
\hat{R}_{-} \hat{S}_{L}^{(j)} \hat{R}_{+} = \sum
\limits_{L'=0}^{2j} h_{L'}\hat{S}_{L'}^{(j)},
\end{equation}

\noindent with

\begin{eqnarray}
h_{L'} && = \left[ {\rm Tr} \left( \left. \hat{S}_{L'}^{(j)}
\right. ^{2} \right) \right]^{-1} \left[ {\rm Tr} \left(
\hat{S}_{L}^{(j)}\hat{S}_{L'}^{(j)} \right) -\frac{1}{2}{\rm Tr}
\left( \hat{R}_{+}\hat{S}_{L}^{(j)}\hat{R}_{-}\hat{S}_{L'}^{(j)}
\right) -\frac{1}{2}{\rm Tr} \left(
\hat{R}_{+}\hat{S}_{L'}^{(j)}\hat{R}_{-}\hat{S}_{L}^{(j)}
\right)\right] \nonumber\\
&& = \delta_{LL'} \left\{ 1 - {\rm Tr} \left(
\hat{R}_{+}\hat{T}_{L0}^{(j)}\hat{R}_{-}\hat{T}_{L0}^{(j)} \right)
\right\}.
\end{eqnarray}

\noindent Now, in view of (\ref{TL0}) and (\ref{R+R-::matrices}),
we obtain

\begin{equation}
{\rm Tr} \left(
\hat{R}_{+}\hat{T}_{L0}^{(j)}\hat{R}_{-}\hat{T}_{L0}^{(j)} \right)
= - \sum \limits_{m=-j}^{j-1} \langle jm; j-m | L0 \rangle \langle
j,m+1 ; j,-m-1 | L0 \rangle = \frac{2(j-L)}{2j+1}
\end{equation}

\noindent and

\begin{equation}
\label{h:LL':result} h_{L'} = \delta_{LL'} \frac{2L+1}{2j+1}.
\end{equation}

\noindent Substituting (\ref{h:LL':result}) for $h_{L'}$ in
(\ref{S-RSR-RSR::through:h:S}) and combining the result obtained
with (\ref{uDu::complex:S}), we get formula
(\ref{quantizer:expansion:S}). This completes the proof that, for
the quantizer operator, exponential expression
(\ref{quantizer:exp}) is equivalent to expansion
(\ref{quantizer:expansion:S}) through orthogonal operators
$\hat{S}_{L}^{(j)}$ and, consequently, to formula
(\ref{quantizer:complex}) expressed in terms of irreducible tensor
operators.

\section{\label{section:delta} Kernel of the delta-function on the tomogram set}

Using definitions (\ref{tomogram}) and (\ref{rho:reconstruction}),
one can easily write

\begin{equation}
w({\bf x}_{1}) = \int w({\bf x}_{2}) {\rm Tr}\left( \hat{D}({\bf
x}_{2}) \hat{U}({\bf x}_{1}) \right) {\rm d}{\bf x}_{2}.
\end{equation}

\noindent This implies that the function ${\rm Tr}\left(
\hat{D}({\bf x}_{2}) \hat{U}({\bf x}_{1}) \right)$ can be treated
as the kernel of the unity operator on the set of spin tomograms.
As far as qubits are considered, we employ the exact formulas for
quantizer and dequantizer operators
(\ref{U:qubit:explicitmatrix}). The result is

\begin{equation}
\label{delta:qubit} {\rm Tr}\left( \hat{D}({\bf x}_{2})
\hat{U}({\bf x}_{1}) \right) = \frac{1}{2} + 6 m_{1} m_{2} \left(
\cos\beta_{1} \cos\beta_{2} + \sin\beta_{1} \sin\beta_{2}
\cos(\alpha_{1}-\alpha_{2}) \right) = \frac{1}{2} + 6 m_{1} m_{2}
\left( {\bf n}_{1} \cdot {\bf n}_{2} \right).
\end{equation}

\noindent In case of qutrits, analogues calculations, with account
of (\ref{U:qutrit:explicitmatrix}), yield

\begin{equation}
\label{delta:qutrit} {\rm Tr}\left( \hat{D}({\bf x}_{2})
\hat{U}({\bf x}_{1}) \right) = \frac{1}{3} + \frac{3}{2} m_{1}
m_{2} \left( {\bf n}_{1} \cdot {\bf n}_{2} \right) + \frac{5}{12}
(3m_{1}^{2}-2)(3m_{2}^{2}-2) \left( 3({\bf n}_{1} \cdot {\bf
n}_{2})^{2} - 1 \right).
\end{equation}

\noindent Here we introduced vectors ${\bf n}_{i}$, which
correspond to matrices $u(\alpha_{i},\beta_{i},\gamma_{i})$
according to the rule

\begin{equation}
{\bf n}_{i} = (\cos\alpha_{i}\sin\beta_{i},
\sin\alpha_{i}\sin\beta_{i}, \cos\beta_{i})
\end{equation}

\noindent and determine the axis of quantization of the spin
projection for the operator $u^{\dag}\hat{J}_{z}u$.

\section{\label{section:starproduct} Star-product for qubit and qutrit tomograms}

By construction, the tomographic symbol $f_{\hat{A}}({\bf x})$ is
related to the operator $\hat{A}$ as follows:

\begin{equation}
f_{\hat{A}}({\bf x}) = {\rm Tr} \left( \hat{A}\hat{U}({\bf x})
\right), \qquad \hat{A} = \int f_{\hat{A}}({\bf x}) \hat{D}({\bf
x}) {\rm d}{\bf x}.
\end{equation}

\noindent The symbol of the operator $\hat{A}\hat{B}$ is called
the star-product of symbols $f_{\hat{A}}({\bf x})$ and
$f_{\hat{B}}({\bf x})$. In other words,

\begin{equation}
\label{star:product:scheme} f_{\hat{A}\hat{B}}({\bf x}_{1}) =
f_{\hat{A}}({\bf x}_{1}) \ast f_{\hat{B}}({\bf x}_{1}) = {\rm Tr}
\left( \hat{A}\hat{B}\hat{U}({\bf x}_{1}) \right) = \int\!\!\!\int
f_{\hat{A}}({\bf x}_{3}) f_{\hat{B}}({\bf x}_{2}) K({\bf
x}_{3},{\bf x}_{2},{\bf x}_{1}) {\rm d}{\bf x}_{2} {\rm d}{\bf
x}_{3},
\end{equation}

\noindent where the function

\begin{equation}
\label{kernel:general} K({\bf x}_{3},{\bf x}_{2},{\bf x}_{1}) =
{\rm Tr} \left( \hat{D}({\bf x}_{3})\hat{D}({\bf
x}_{2})\hat{U}({\bf x}_{1}) \right)
\end{equation}

\noindent is called the kernel of the star-product scheme.

Direct calculations of kernel (\ref{kernel:general}) for qubit
case yield

\begin{eqnarray}
\label{kernel:qubit} K({\bf x}_{3},{\bf x}_{2},{\bf x}_{1}) = &&
\frac{1}{4} + 3m_{1}m_{2} \left( {\bf n}_{1} \cdot {\bf n}_{2}
\right) + 9m_{2}m_{3} \left( {\bf n}_{2} \cdot {\bf n}_{3} \right)
+ 3m_{1}m_{3} \left( {\bf
n}_{3} \cdot {\bf n}_{1} \right) \nonumber\\
&&+ i 18 m_{1}m_{2}m_{3} \left( {\bf n}_{1} \cdot [{\bf n}_{2}
\times {\bf n}_{3}] \right).
\end{eqnarray}

\noindent Here, $\left( {\bf n}_{1} \cdot [{\bf n}_{2} \times {\bf
n}_{3}] \right)$ denotes the scalar triple product of vectors
${\bf n}_{1}$, ${\bf n}_{2}$, and ${\bf n}_{3}$.

As far as qutrits are concerned, kernel (\ref{kernel:general})
takes the form

\begin{eqnarray}
\label{kernel:qutrit}
 K({\bf x}_{3},{\bf x}_{2},{\bf x}_{1}) = && \frac{1}{9} +
\frac{1}{2}m_{1}m_{2} \left( {\bf n}_{1} \cdot {\bf n}_{2} \right)
+ \frac{3}{2} m_{2}m_{3} \left( {\bf n}_{2} \cdot {\bf n}_{3}
\right) + \frac{1}{2}m_{1}m_{3} \left( {\bf n}_{3} \cdot {\bf
n}_{1} \right) \nonumber\\
&&+ i\frac{9}{8} m_{1}m_{2}m_{3} \left( {\bf n}_{1} \cdot [ {\bf
n}_{2} \times {\bf n}_{3} ] \right) \nonumber\\
&& + \frac{5}{36}(3m_{1}^{2}-2)(3m_{2}^{2}-2) \left( 3({\bf n}_{1}
\cdot {\bf n}_{2})^{2} - 1 \right)   \nonumber\\
&&+ \frac{25}{36}(3m_{2}^{2}-2)(3m_{3}^{2}-2) \left( 3({\bf n}_{2}
\cdot {\bf n}_{3})^{2} - 1 \right) \nonumber\\
&&+ \frac{5}{36}(3m_{1}^{2}-2)(3m_{3}^{2}-2) \left( 3({\bf n}_{3}
\cdot {\bf n}_{1})^{2} - 1 \right)  \nonumber\\
&&+ \frac{3}{8}(3m_{1}^{2}-2)m_{2}m_{3} \Big( 3({\bf n}_{1} \cdot
{\bf n}_{2})({\bf n}_{1} \cdot {\bf n}_{3}) - ({\bf n}_{2} \cdot
{\bf n}_{3}) \Big)  \nonumber\\
&& + \frac{5}{8}m_{1}(3m_{2}^{2}-2)m_{3} \Big( 3({\bf n}_{2} \cdot
{\bf n}_{3})({\bf n}_{2} \cdot {\bf n}_{1}) - ({\bf n}_{3} \cdot
{\bf n}_{1}) \Big) \nonumber\\
&&+\frac{5}{8}m_{1}m_{2}(3m_{3}^{2}-2)
\Big( 3({\bf n}_{3} \cdot {\bf n}_{1})({\bf n}_{3} \cdot {\bf n}_{2}) - ({\bf n}_{1} \cdot {\bf n}_{2}) \Big) \nonumber\\
&& + i\frac{25}{8} m_{1}(3m_{2}^{2}-2)(3m_{3}^{2}-2) \left( {\bf
n}_{2} \cdot {\bf n}_{3} \right) \left( {\bf n}_{1} \cdot [{\bf
n}_{2} \times {\bf n}_{3}] \right)  \nonumber\\
&&+ i\frac{15}{8} (3m_{1}^{2}-2)m_{2}(3m_{3}^{2}-2) \left( {\bf
n}_{3} \cdot {\bf n}_{1} \right) \left( {\bf n}_{1} \cdot [{\bf
n}_{2} \times {\bf n}_{3}] \right) \nonumber\\
&& + i\frac{15}{8} (3m_{1}^{2}-2)(3m_{2}^{2}-2)m_{3} \left( {\bf
n}_{1} \cdot {\bf n}_{2} \right) \left( {\bf n}_{1} \cdot [{\bf
n}_{2} \times {\bf n}_{3}] \right)  \nonumber\\
&&+ \frac{25}{72} (3m_{1}^{2}-2)(3m_{2}^{2}-2)(3m_{3}^{2}-2) \nonumber\\
&& \qquad \qquad \qquad \qquad \times \Big\{ 3 \left( {\bf n}_{1}
\cdot {\bf n}_{2} \right) \Big( \left[{\bf n}_{1} \times {\bf
n}_{3}\right] \cdot \left[{\bf n}_{2} \times {\bf n}_{3}\right]
\Big) \nonumber\\
&& \qquad \qquad \qquad \qquad \qquad + 3 \left( {\bf n}_{2} \cdot
{\bf n}_{3} \right) \Big( \left[{\bf n}_{2} \times {\bf
n}_{1}\right] \cdot \left[{\bf
n}_{3} \times {\bf n}_{1}\right] \Big) \nonumber\\
&& \qquad \qquad \qquad \qquad \qquad + 3 \left( {\bf n}_{3} \cdot
{\bf n}_{1} \right) \Big( \left[{\bf n}_{3} \times {\bf
n}_{2}\right] \cdot \left[{\bf n}_{1} \times {\bf n}_{2}\right]
\Big) - 2 \Big\},
\end{eqnarray}

\noindent where $\left[ {\bf n}_{i} \times {\bf n}_{j} \right]$ is
the cross product of vectors ${\bf n}_{i}$ and ${\bf n}_{j}$.

From (\ref{star:product:scheme}) it follows that, if
$\hat{A}=\hat{1}$, then

\begin{equation}
f_{\hat{B}}({\bf x}_{1}) = \int f_{\hat{B}}({\bf x}_{2}) \left(
\int K({\bf x}_{3},{\bf x}_{2},{\bf x}_{1}) {\rm d}{\bf x}_{3}
\right) {\rm d}{\bf x}_{2}.
\end{equation}

\noindent This implies that

\begin{equation}
\label{delta-kernel-relation} {\rm Tr}\left( \hat{D}({\bf x}_{2})
\hat{U}({\bf x}_{1}) \right) = \int K({\bf x}_{3},{\bf x}_{2},{\bf
x}_{1}) {\rm d}{\bf x}_{3}.
\end{equation}

\noindent Employing explicit formulas (\ref{delta:qubit}),
(\ref{delta:qutrit}), (\ref{kernel:qubit}), and
(\ref{kernel:qutrit}), one can easily check that requirement
(\ref{delta-kernel-relation}) is satisfied for qubits and qutrits.

\section{\label{section:dual:symbols} Dual tomographic symbols}

Dual tomographic symbols are especially convenient for calculating
the expectation values of observables, i.e., the quantity ${\rm
Tr}(\hat{\rho}\hat{A})$. Indeed, the trace of the product of two
operators $\hat{A}$ and $\hat{B}$ reads

\begin{equation}
{\rm Tr} \left(\hat{A}\hat{B}\right) = \int f_{\hat{A}}({\bf x})
{\rm Tr} \left( \hat{B}\hat{D}({\bf x}) \right) {\rm d}{\bf x} =
\int f_{\hat{A}}({\bf x}) f_{\hat{B}}^{d}({\bf x}) {\rm d}{\bf x},
\end{equation}

\noindent where we introduced the dual tomographic symbol of the
operator $\hat{B}$ as follows:

\begin{equation}
f_{\hat{B}}^{d}({\bf x}) = {\rm Tr} \left( \hat{B}\hat{D}({\bf x})
\right), \qquad \hat{B} = \int f_{\hat{B}}^{d}({\bf x})
\hat{U}({\bf x}) {\rm d}{\bf x}.
\end{equation}

It is easy to prove that the star-product kernel for dual
tomographic symbols takes the form

\begin{equation}
\label{kernel:dual:general} K^{d}({\bf x}_{3},{\bf x}_{2},{\bf
x}_{1}) = {\rm Tr} \left( \hat{U}({\bf x}_{3})\hat{U}({\bf
x}_{2})\hat{D}({\bf x}_{1}) \right).
\end{equation}

\noindent Due to the similar structure of quantizer and
dequantizer operators [see Eqs. (\ref{dequantizer:expansion:S})
and (\ref{quantizer:expansion:S})], the kernel $K^{d}({\bf
x}_{3},{\bf x}_{2},{\bf x}_{1})$ differs from the kernel $K({\bf
x}_{3},{\bf x}_{2},{\bf x}_{1})$ by numerical factors of the
corresponding terms.

\noindent For qubits, one has

\begin{eqnarray}
K^{d}({\bf x}_{3},{\bf x}_{2},{\bf x}_{1}) = && \frac{1}{4} +
3m_{1}m_{2} \left( {\bf n}_{1} \cdot {\bf n}_{2} \right) +
m_{2}m_{3} \left( {\bf n}_{2} \cdot {\bf n}_{3} \right) +
3m_{1}m_{3} \left( {\bf
n}_{3} \cdot {\bf n}_{1} \right) \nonumber\\
&&+ i 6 m_{1}m_{2}m_{3} \left( {\bf n}_{1} \cdot [{\bf n}_{2}
\times {\bf n}_{3}] \right).
\end{eqnarray}

\noindent In case of qutrits, we obtain

\begin{eqnarray}
 K^{d}({\bf x}_{3},{\bf x}_{2},{\bf x}_{1}) = && \frac{1}{9} +
\frac{1}{2}m_{1}m_{2} \left( {\bf n}_{1} \cdot {\bf n}_{2} \right)
+ \frac{1}{6} m_{2}m_{3} \left( {\bf n}_{2} \cdot {\bf n}_{3}
\right) + \frac{1}{2}m_{1}m_{3} \left( {\bf n}_{3} \cdot {\bf
n}_{1} \right) \nonumber\\
&&+ i\frac{3}{8} m_{1}m_{2}m_{3} \left( {\bf n}_{1} \cdot [ {\bf
n}_{2} \times {\bf n}_{3} ] \right) \nonumber\\
&& + \frac{5}{36}(3m_{1}^{2}-2)(3m_{2}^{2}-2) \left( 3({\bf n}_{1}
\cdot {\bf n}_{2})^{2} - 1 \right)   \nonumber\\
&&+ \frac{1}{36}(3m_{2}^{2}-2)(3m_{3}^{2}-2) \left( 3({\bf n}_{2}
\cdot {\bf n}_{3})^{2} - 1 \right) \nonumber\\
&&+ \frac{5}{36}(3m_{1}^{2}-2)(3m_{3}^{2}-2) \left( 3({\bf n}_{3}
\cdot {\bf n}_{1})^{2} - 1 \right)  \nonumber\\
&&+ \frac{5}{24}(3m_{1}^{2}-2)m_{2}m_{3} \Big( 3({\bf n}_{1} \cdot
{\bf n}_{2})({\bf n}_{1} \cdot {\bf n}_{3}) - ({\bf n}_{2} \cdot
{\bf n}_{3}) \Big)  \nonumber\\
&& + \frac{1}{8}m_{1}(3m_{2}^{2}-2)m_{3} \Big( 3({\bf n}_{2} \cdot
{\bf n}_{3})({\bf n}_{2} \cdot {\bf n}_{1}) - ({\bf n}_{3} \cdot
{\bf n}_{1}) \Big)  \nonumber\\
&&+ \frac{1}{8}m_{1}m_{2}(3m_{3}^{2}-2)
\Big( 3({\bf n}_{3} \cdot {\bf n}_{1})({\bf n}_{3} \cdot {\bf n}_{2}) - ({\bf n}_{1} \cdot {\bf n}_{2}) \Big) \nonumber\\
&& + i\frac{3}{8} m_{1}(3m_{2}^{2}-2)(3m_{3}^{2}-2) \left( {\bf
n}_{2} \cdot {\bf n}_{3} \right) \left( {\bf n}_{1} \cdot [{\bf
n}_{2} \times {\bf n}_{3}] \right)  \nonumber\\
&&+ i\frac{5}{8} (3m_{1}^{2}-2)m_{2}(3m_{3}^{2}-2) \left( {\bf
n}_{3} \cdot {\bf n}_{1} \right) \left( {\bf n}_{1} \cdot [{\bf
n}_{2} \times {\bf n}_{3}] \right) \nonumber\\
&& + i\frac{5}{8} (3m_{1}^{2}-2)(3m_{2}^{2}-2)m_{3} \left( {\bf
n}_{1} \cdot {\bf n}_{2} \right) \left( {\bf n}_{1} \cdot [{\bf
n}_{2} \times {\bf n}_{3}] \right)  \nonumber\\
&&+ \frac{5}{72} (3m_{1}^{2}-2)(3m_{2}^{2}-2)(3m_{3}^{2}-2) \nonumber\\
&& \qquad \qquad \qquad \qquad \times \Big\{ 3 \left( {\bf n}_{1}
\cdot {\bf n}_{2} \right) \Big( \left[{\bf n}_{1} \times {\bf
n}_{3}\right] \cdot \left[{\bf n}_{2} \times {\bf n}_{3}\right]
\Big) \nonumber\\
&& \qquad \qquad \qquad \qquad \qquad + 3 \left( {\bf n}_{2} \cdot
{\bf n}_{3} \right) \Big( \left[{\bf n}_{2} \times {\bf
n}_{1}\right] \cdot \left[{\bf
n}_{3} \times {\bf n}_{1}\right] \Big) \nonumber\\
&& \qquad \qquad \qquad \qquad \qquad + 3 \left( {\bf n}_{3} \cdot
{\bf n}_{1} \right) \Big( \left[{\bf n}_{3} \times {\bf
n}_{2}\right] \cdot \left[{\bf n}_{1} \times {\bf n}_{2}\right]
\Big) - 2 \Big\}.
\end{eqnarray}

\section{\label{section:conclusions} Conclusions}

The tomographic-probability representation of quantum mechanics
allows one to describe states and operators by special functions
(tomographic symbols). Moreover, the tomograms can be measured
experimentally. Spin tomography has undergone fast development in
the last few decades and has been attacked with the help of
different approaches. We managed here to demonstrate the
equivalency of two methods available in the literature. We also
succeeded in developing a simple form of dequantizer and quantizer
operators needed for scanning and reconstruction procedure,
respectively. The explicit form of the star-product kernel is
obtained for qubits and qutrits. Utilizing these expressions is
straightforward while we deal with ordinary or dual tomographic
symbols of operators.

\section*{Acknowledgments}
V.I.M. thanks the Russian Foundation for Basic Research for
partial support under Project Nos. 07-02-00598 and 08-02-90300.
S.N.F. thanks the Ministry of Education and Science of the Russian
Federation and the Federal Education Agency for support under
Project No. 2.1.1/5909.

\end{document}